\documentclass[12pt,a4paper,final]{iopart}

\usepackage{iopams}
\usepackage{graphicx}
\usepackage[breaklinks=true,colorlinks=true,linkcolor=blue,urlcolor=blue,citecolor=blue]
{hyperref}
\usepackage[latin1]{inputenc}
\usepackage{bm}
\usepackage{epsfig}
\usepackage{amsthm}
\usepackage{dcolumn}
\usepackage{graphicx}
\usepackage{amsfonts}
\usepackage{float}
\expandafter\let\csname equation*\endcsname\relax
\expandafter\let\csname endequation*\endcsname\relax
\usepackage{amsmath,amssymb}
\usepackage{ulem,color,graphics}
\def\bit{\begin{itemize}}
\def\eit{\end{itemize}}
\def\bnu{\begin{enumerate}}
\def\enu{\end{enumerate}}

\def\x{\times}
\def\x{\times}
\def\xb {{\bf x}}
\def\yb {{\bf y}}

\def\rb {{\bf r}}
\def\pb {{\bf p}}

\def\pb {{\bf p}}
\def\qb {{\bf q}}
\def\nn{\nonumber}

\def\nn{\nonumber}
\def\nn{\nonumber}

\def\ie{{\em i.e.,}}

\def\ie{{\it i.e.,}}
\def\go{\rightarrow}
\def\go{\rightarrow}
\def\go{\rightarrow}
\def\fot{\frac{1}{2}}
\def\rh {\hat{\bf r}}
\def\ph {\hat{\bf p}}

\def\xh {\hat{\bf x}}
\def\yh {\hat{\bf y}}
\def\Mass{\mathrm{M}}

\def\eit{\end{itemize}}
\def\bra#1{\langle #1|}
\def\ket#1{|#1 \rangle}
\def\rf#1{{(\ref{#1})}}
\def\rf#1{{(\ref{#1})}}
\def\bra#1{\langle #1|}
\def\ket#1{|#1 \rangle}
\def\bra#1{\langle #1|}
\def\ket#1{|#1 \rangle}
\def\rf#1{{(\ref{#1})}}
\def\etal{{\it et al.}}
\def\Ket#1{||#1 \rangle}
\def\Bra#1{\langle #1||}
\def\Ket#1{||#1 \rangle}
\def\Bra#1{\langle #1||}
\def\Ket#1{||#1 \rangle}
\def\Bra#1{\langle #1||}
\def\go{\rightarrow}

\def\Psip{^\uparrow\Psi}
\def\psip{^\uparrow\psi}
\def\bit{\begin{itemize}}

\def\psin{^\downarrow\psi}
\def\psin{^\downarrow\psi}
\def\Psin{^\downarrow\Psi}
\def\beq{\begin{equation}}
\def\eeq{\end{equation}}
\def\bea{\begin{eqnarray}}
\def\eea{\end{eqnarray}}
\def\be{\begin{equation}}
\def\ee{\end{equation}}
\def\br{\begin{eqnarray}}
\def\er{\end{eqnarray}}
\def\brn{\begin{eqnarray*}}
\def\ern{\end{eqnarray*}}

\def\ov#1#2{\langle #1 | #2  \rangle}
\def\ov#1#2{\langle #1 | #2  \rangle}
\def\ov#1#2{\langle #1 | #2  \rangle }

\def\bin#1#2{\left(\negthinspace\begin{array}{c}#1\\#2\end{array}\right)}
\def\bin#1#2{\left(\negthinspace\begin{array}{c}#1\\#2\end{array}\right)}
\def\cbin#1#2{\left(\negthinspace\begin{array}{cc}#1&#2\end{array}\right)}

\def\sixj#1#2#3#4#5#6{\left\{\negthinspace\begin{array}
{ccc}#1&#2&#3\\#4&#5&#6\end{array}\right\}}
\def\threej#1#2#3#4#5#6{\left(\negthinspace\begin{array}
{ccc}#1&#2&#3\\#4&#5&#6\end{array}\right)}

\def\M {{{\cal M}}}
\def\A {{{\cal A}}}
\def\B {{{\cal B}}}

\newcommand{\Ref}[1]{{(\ref{#1})}}

\newcommand{\sDelta}{\mathsf{\Delta}}
\newcommand{\sgn}{\mathrm{sgn}}

\begin{document}
\title{Nonmesonic weak decay of charmed hypernuclei}

\author{C E Fontoura$^1$, F Krmpoti\'c$^{2}$, A P Gale\~ao$^{3}$, C De Conti$^{4}$ 
and \\ G Krein$^{3}$}

\address{$^1$Instituto Tecnol\'ogico de Aeron\'autica, DCTA, 12228-900 S\~ao
Jos\'e dos Campos, SP, Brazil\\
$^2$Instituto de F\'isica La Plata, Universidad Nacional de La Plata \\
1900 La Plata, Argentina \\
$^3$Instituto de F\'isica Te\'orica, Universidade Estadual Paulista \\
Rua Dr. Bento Teobaldo Ferraz, 271 - Bloco II, 01140-070, 
S\~ao Paulo, SP, Brazil
\\
$^4$Campus Experimental de Rosana, Universidade Estadual Paulista \\
19274-000 Rosana, SP, Brazil}

\vspace{10pt}

\begin{abstract}
We present a study of the nonmesonic weak decay (NMWD) of charmed hypernuclei using 
a relativistic formalism. We work within the framework of the independent particle shell 
model and employ a ($\pi$,$K$) one-meson-exchange model for the decay dynamics. 
We implement a fully relativistic treatment of nuclear recoil. Numerical results are
obtained for the one-neutron-induced transition NMWD rates of the $_{\Lambda^{+}_{c}}^{12}$N. 
The effect of nuclear recoil is sizable and goes in the direction to decrease the nuclear 
decay rate. We found that the NMWD decay rate of $_{\Lambda^{+}_{c}}^{12}$N is of the same order 
of magnitude as the partial decay rate for the corresponding mesonic decay $\Lambda^+_c \rightarrow 
\Lambda + \pi^+$, suggesting the feasibility of experimental detection of such heavy-flavor 
nuclear processes. 
\end{abstract}


\vspace{2pc}

\section{Introduction}
\label{sec:intro}

The field of hadron spectroscopy has been galvanized by the continuous discovery of the 
so-called X,Y,Z exotic hadrons since the discovery by the Belle collaboration~\cite{Choi:2003ue}
in 2003 of the charmed hadron $X(3872)$. They are {\it exotic} because they do not fit the 
conventional quark-model pattern of either quark-antiquark mesons or three-quark baryons. Most of 
the X,Y,Z hadrons have masses close to open heavy-flavor thresholds and decay into hadrons containing 
charm (or bottom) quarks~\cite{Lebed:2016hpi}.  On a parallel route in nuclear physics, there has 
been growing interest in the study of the interactions of charmed hadrons with atomic 
nuclei~\cite{{Krein:2016fqh},{Briceno:2015rlt},{Hosaka:2016ypm}}. Several investigations have
predicted the existence of nuclear bound states with charmed mesons~\cite{Dmesic-Tsu,
Yasui:2009bz,GarciaRecio:2010vt,GarciaRecio:2011xt,Krein:2010vp,Tsushima:2011kh,Dmesic-Tol} 
and charmed baryons~\cite{Tsushima:2002ua,Tsushima:2003dd,Gar15,Maeda:2015hxa,{Shyam:2016uxa}}.
The study of such systems is of great scientific interest since new degrees of freedom 
are introduced into the traditional world of nuclei by revealing the existence
of new forms of nuclear matter. 

Historically, single-$\Lambda$ hypernuclei (with strangeness $S=-1$) represent the first kind 
of flavored nuclei with nonzero strangeness ever observed~\cite{first}, an event that marks 
the inauguration of a new branch of nuclear physics, hypernuclear physics. The field has developed 
in an independent direction {\textemdash} Refs.~\cite{Feliciello:2015dua} and \cite{Gal16} are 
recent reviews on experiment and theory, respectively. Presently different kinds of hypernuclei as
doubly-strange hypernuclei~\cite{CBM,{Nakazawa:2017oac},{Nagae:2017slp},{Kanatsuki:2017iew}}, 
antihypertriton~\cite{star} and exotic hypernuclei~\cite{GSI1,GSI2,GSI3} are vigorously studied.

The possibility to form $\Lambda^{+}_{c}$ and $\Sigma^{+}_{c}$ hypernuclei was first 
suggested about 40 years ago~\cite{Tyapkin}, soon after the discovery of the charm quark, and a 
first calculation of their binding energies~\cite{Dov77} was performed in the framework of a meson 
exchange model with coupling constants determined by $SU(4)$ flavor symmetry. 
Although the existence of charmed nuclei has not been experimentally demonstrated in a 
conclusive way~\cite{exp1,exp2}, several authors in the succeeding decades have found, using 
different models for the interactions between nucleons and charmed baryons, that 
such hypothetical flavored nuclei could actually form a rich spectrum of bound 
systems~\cite{Iwao,{Gibson:1983zw},Tsushima:2002ua,Tsushima:2003dd}. The experimental situation 
can change in a few years, with the starting of operation of the FAIR facility in Germany, 
and the extension of the Hadron Hall at the JPARC Laboratory in Japan, where the present 
proton beam will be used by adding in the extension a secondary target to produce antiprotons 
for charmed hadron production. 

Before approaching the weak decay of $\Lambda^{+}_{c}$ hypernuclei, let us recall 
some well known facts about that of $\Lambda$ hypernuclei.
The free $\Lambda$ hyperon decays mainly via the pionic modes~\cite{PDG}
\br
\Lambda&\go& p+\pi^++38~{\rm  MeV}~ (64\%),
\nn\\
\Lambda
&\go& n+\pi^0+41~{\rm  MeV}~ (36\%),
\label{1}
\er
with a lifetime of $\tau_{\Lambda} = (2.63 \pm 0.02) \x 10^{-10}$ s.
These same decay modes take place within a  $\Lambda$ hypernucleus, but the 
$\Lambda$ hyperon  is now bound and the energy of the released  nucleon $N=p,n$  
is small ($\leq 10$ MeV) in comparison with the Fermi energy 
$\epsilon_F=37$ MeV. Thus, the pionic decay modes are severely inhibited by Pauli blocking of the 
final-state nucleons, which makes the hypernuclear mesonic decay rate $\Gamma_m$ 
to be relatively small compared with the free decay rate, $\Gamma_{\Lambda} = \hbar/\tau_{\Lambda} 
= (2.50 \pm 0.02) \x 10^{-6}$ MeV, in all but the lightest hypernuclei. This fact potentiates the 
occurrence of the nonmesonic weak decay (NMWD) reaction 
\br
\Lambda+N\rightarrow N+N,
\label{2}
\er
within the hypernucleus, which liberates enough kinetic energy to put the two emitted nucleons 
above the Fermi surface. As a consequence, the NMWD dominates over the mesonic mode in medium and 
heavy hypernuclei and has a decay rate $\Gamma_{nm}\equiv\Gamma_{p}+\Gamma_{n}$ which is about 
of the same value as~$\Gamma_{\Lambda}$ {\textemdash} it must be mentioned that there is also a 
somewhat sizable contribution from two-nucleon induced channels to 
$\Gamma_{nm}$~\cite{{Feliciello:2015dua},Gal16}, which we are not taken into account here.
Needless to say that the investigation of the dynamics of the NMWD in $\Lambda$ hypernuclei is an 
indispensable tool to inquire about the baryon-baryon strangeness-changing interaction, and many 
experimental~\cite{Buf13,Ag14} and theoretical~\cite{{McKGib},Du96,Pa97,It02,Ba03,Ba05,
Garbarino:2013rwa,{Chumillas},{Cheung},{Heddle},{Inoue},{Sasaki1},{Sasaki2},{Bauer2},{Bauer3}} groups 
have concentrated efforts on this subject {\textemdash} for a more complete 
list of references, see e.g. the review in Ref.~\cite{Botta}. Recently, it was suggested that the $\pi+K$ 
meson-exchange model with soft monopole form factors could be a good starting point to describe this type 
of interaction in light and medium systems~\cite{Kr14,Kr14a}.

The lifetime of  the free charmed baryon $\Lambda^{+}_c$ is 
$\tau_{\Lambda^{+}_{c}} = (2.00 \pm 0.06) \x 10^{-13}$ s,  which corresponds to the decay rate
$\Gamma(\Lambda^{+}_c\rightarrow all) = (3.29 \pm 0.10) \x 10^{-9}$ MeV.  Among several hadronic 
decay channels with a hyperon in the final state, it also decays via the pionic 
mode~\cite{PDG}
\br
\Lambda^{+}_c&\go& \Lambda+\pi^++1030~{\rm  MeV}~ (1.07\%),
\label{3}
\er
with a partial width of 
$\Gamma(\Lambda^{+}_c\rightarrow\Lambda +\pi) = (4.28 \pm 0.26) \x 10^{-11}$ MeV.
This mesonic decay  can also take place inside a  $\Lambda^{+}_c$ charmed-nucleus, and 
since the produced $\Lambda$ hyperon is not Pauli blocked, it will have a decay rate similar 
to that of the free $\Lambda^{+}_c$~\cite{Ghosh:2016jjv}.

Bunyatov \etal ~\cite{Bunyatov:1992in} have suggested long ago that $\Lambda^+_c$ hypernuclei,
analogously to $\Lambda$ hypernuclei, may also decay nonmesonically. In this case, the NMWD 
reaction would be driven by the reaction
\br
\Lambda^{+}_c+n\rightarrow\Lambda+p.
\label{4}
\er
However, there are important differences between the NMWD of 
$\Lambda$ hypernuclei and that of $\Lambda^+_c$ hypernuclei. One
of the most important differences concerns the energy liberated in the charmed hyperon 
decay ($\simeq \Delta_c=M_{\Lambda^{+}_{c}}-M_{\Lambda} =1.170$~GeV), which is several times bigger 
than that of the strange hyperon ($\simeq\Delta_s= M_{\Lambda}-M_{N} =177$ MeV). A consequence 
of this large energy release is that nonrelativistic approaches might become inapplicable for the 
evaluation of NMWD transition  matrix elements in charmed nuclei. In addition, a large energy
release also implies that nuclear recoil cannot be neglected in the calculation of decay
rates, particularly for light- and medium-weight nuclei.  It is therefore important to examine 
the impact of relativistic effects on the decay rates. On the other hand, the interactions of 
the fast outgoing baryons with the residual nuclear system are expected to play a minor role.

In the present work we use the relativistic formalism developed in Ref.~\cite{Fontoura:2015kta} 
for the NMWD of $\Lambda$ hypernuclei to investigate the similar decay process in 
$\Lambda^+_c$ hypernuclei. The formalism is based on an independent-particle shell model. The
application of a relativistic model for the study of the structure of hypernuclei
dates back to the late 1970's~\cite{BroWei}, but so far not much is known about
the impact of a relativistic approach in the evaluation of NMWD rates. The first
studies started two decades ago~\cite{Ra91} using single-particle bound-state
wave functions obtained by solving the Dirac equation with static Lorentz-scalar 
and Lorentz-vector Woods-Saxon potentials, and transition matrix elements calculated with 
the pseudoscalar ($\pi$,$K$) one-meson-exchange model. A similar relativistic approach for 
the nuclear structure was described in Refs.~\cite{Co09,Co09a}. 

We implement a fully relativistic treatment of recoil. Short-range correlations in
the initial state, that arise due to the overlap of the wave functions of $\Lambda^+_c$ and nucleons 
in the hypernucleus are not captured in a mean field treatment of nuclear structure, 
but  are expected to be of less importance in the NMWD of a charmed hypernucleus than of a strange 
hypernucleus. This is because the short-range repulsion in $\Lambda^+_c-N$ is much weaker than in $\Lambda-N$, as 
indicated by a recent lattice QCD calculation~\cite{Miyamoto:2016hqo}. Therefore, in this first study 
we neglect their effects in the calculation of decay rates.

The paper is organized as follows. In Section~\ref{IPSM} we explain the
general shell model formalism for the NMWD of single $\Lambda^+_c$ hypernuclei.  Next, 
in Section~\ref{two-body}, we deal with the expression for the two-body NMWD transition amplitude.
Then, in Section~\ref{rate}, we discuss the calculation scheme to obtain the 
decay rate for charmed nuclei with open- and closed-shell cores, at first without taking 
recoil effects into account. Subsequently, in Section~\ref{recoil}, these effects are 
discussed in the relativistic framework. Numerical results for the NMWD  of 
$_{\Lambda^{+}_{c}}^{12}$N are presented  in Section~\ref{results}, where we also examine 
the impact of the fully relativistic treatment of the recoil effect on the decay 
rate and some related spectral distributions.  Our conclusions and perspectives for future work are 
presented in Section~\ref{sec7}. The paper contains three appendices; in \ref{A} we present details on
the bound and continuum single-particle wave functions used in the calculation of the decay rates.
We also present numerical results for the single-particle energies. \ref{B} implements the
the partial-wave decomposition of the decay amplitude. Finally, \ref{C} collects details on the
integration over the outgoing proton when using the relativistic formalism of nuclear recoil.


\section{Relativistic independent-particle shell model}
\label{IPSM}

The nuclear structure aspects of the charmed NMWD will be described 
in the framework of a relativistic version the spherical Independent Particle 
Shell Model (IPSM). The charmed hypernucleus with $A$ baryons is assumed to be 
in its ground state, which is taken as 
a charmed baryon $\Lambda^{+}_{c}$ in the single-particle state 
$j_{\Lambda^{+}_{c}}=1s_{1/2}$ weakly coupled to the appropriate $(A-1)$ nuclear
core of spin $J_C$, 
forming an initial state of spin $J_I$,
\ie 
\be
\ket{J_I} \equiv 
\Big(a^\dagger_{j_{\Lambda^{+}_{c}}}\otimes\ket{J_C}\Big)_{J_I}.
\label{5}
\ee
In the specific case of the hypothetical charmed hypernucleus\footnote{We adhere to 
the notation used in most of the recent~\cite{{Tsushima:2002ua},{Tsushima:2003dd},{Shyam:2016uxa}} 
and past~\cite{{exp1},Gibson:1983zw} literature on charmed hypernuclei, in that a charmed 
$\Lambda^+_c$ hypernucleus with $A$ baryons and total electric charge $Z$ receives the 
name of ordinary nuclides with $Z$ protons. Specifically, in the present case, 
the charmed hypernucleus is composed by $A=12$ baryons and $Z=7$ units of (positive) electric 
charge: one $\Lambda^+_c$, five neutrons and six protons. Therefore, it is denoted 
by $^{12}_{\Lambda^+_c}\mathrm{N}$, where $\mathrm{N}$ stands for Nitrogen. Notice that its nuclear 
structure aspects within the IPSM are analogous to those of the strange hypernucleus $^{12}_\Lambda$C 
dealt with in Ref.~\cite{Fontoura:2015kta}.}\,$_{\Lambda^{+}_{c}}^{12}\mathrm{N}$, the core state
\be
\ket{J_C} = \tilde{a}_{{1p_{3/2}}_n}\,\ket{^{12}\mathrm{C}},
\label{6}
\ee
is assumed to be a $1p_{3/2}$  neutron-hole  with 
respect to the ground state of $^{12}\mathrm{C}$, consisting of completely 
closed $1s_{1/2}$ and $1p_{3/2}$ subshells for both neutrons and protons, which 
is taken to be the Fermi sea. One has to recall that the modified annihilation operators  
$\tilde{a}_{jm} \equiv (-)^{j+m} a_{j\,{-m}}$
are spherical tensors~\cite{Boh69}.
When the neutron inducing the decay is in the single-particle state  
$j_{n}$ ($j\equiv nlj$), the final states of the $(A-2)$ residual nucleus read
\be
\ket{J_F} = \Big( \tilde{a}_{j_{n}}\otimes\ket{J_C}\Big)_{J_F},
\label{7}
\ee
where the final spins  $J_F$ fulfil the constraint $|J_C-j_n|\le J_F\le J_C+j_n$. 
 
The NMWD reaction in Eq.~\rf{4} can be decomposed into transitions in which the two 
initial particles are in intermediate states having total angular momentum $J$.
Doing this, as we shall argue in Section~\ref{rate}, the nuclear structure 
information in the expression for the decay rate will be contained in the 
spectroscopic factors 
\br
F^{j_n}_J &=& \hat{J}^{-2}\sum_{J_F} |\Bra{J_I}
\left( a_{j_n}^\dag a_{j_{\Lambda^{+}_{c}} }^\dag\right)_{J}\Ket{J_F}|^2,
\nn\\
&=&\hat{J}^{2}\sum_{J_F}\sixj{J_C}{J_I}{j_{\Lambda^{+}_{c}}}{J}{j_n}{J_F}^2|
\Bra{J_C}a_{j_n}^\dag\Ket{J_F}|^2.
\label{8}
\er
where we are using the notation $\hat{J}=\sqrt{2J+1}$. The values for $J_I$ and $J_C$ 
are taken from Table I of Ref.~\cite{Kr10}, assuming that they hold also for charmed nuclei,
that is $J_C = 3/2$ and $J_I = 1$. The experimentally measured ground-state spins 
in $^{11}$C and $^{12}_\Lambda$C are, respectively, $3/2^-$ and $1^-$, as can be seen, 
for instance, from Fig. 16 in Ref.~\cite{Gal16}. Regarding $J_I$, the other possible 
value for it would be $J_I=2$. But in the absence of experimental or lattice results on the 
spin-dependent forces in the $\Lambda_c-N$ interaction that ultimately lead to the
splitting between the two states, we have simply assumed for $^{12}_{\Lambda^+_c}$N the measured
value of $J_I$ in $^{12}_\Lambda$C, although a smaller splitting can be expected on the account of 
the larger mass of the $\Lambda^+_c$ that suppresses spin-dependent forces. The corresponding 
values for the factors $F^{j_n}_J$ are listed in Table II of of Ref.~\cite{Kr10}.

The single-particle states for each kind of bound baryon (neutron, proton, 
$\Lambda_c^+$) are the energy eigenfunctions of the respective
single-particle Dirac Hamiltonian 
\be
\hat{h} = -i\bm{\alpha}\cdot\nabla + V_0(r) + \beta[M+S_0(r)],
\label{9}
\ee
where $V_0$ and $S_0$ are spherically-symmetric vector and scalar potentials.
These are constructed in the scheme of the relativistic, spherical, mean-field 
approximation (MFA)~\cite{Ho81,Se86} for the nearest 
doubly-closed-subshell nucleus {\textemdash} see Appendix~B of Ref.~\cite{Fontoura:2015kta}. 
We recall that the evaluation of the matrix elements of the NMWD is made in the 
IPSM and, for consistency, this demands that the $\Lambda^+_c$ wave functions be those
generated by the spherically symmetric mean fields for the $^{12}{\rm C}$ nucleus; 
that is, there is no back reaction of the $\Lambda^+_c$ on the mean fields. 
For nucleons, we choose the potentials corresponding to the model Lagrangian NL3 of 
Ref.~\cite{Lalazissis:1996rd}.
For the $\Lambda_c^+$, they are, in the notation of Ref.~\cite{Fontoura:2015kta}, given by
\br
V_0^{\Lambda_c^+}(r) &=& g_\omega^{\Lambda_c^+} \omega_0(r) + e A_0(r),
\nn\\
S_0^{\Lambda_c^+}(r) &\equiv& S^{\Lambda_c^+}(r) - M_{\Lambda_c^+}
\;=\; g_\sigma^{\Lambda_c^+} \sigma(r).
\label{10}
\er
We use SU(4) flavor symmetry to fix the meson-$\Lambda^+_c$ couplings, 
$g_\omega^{\Lambda_c^+} = g_\omega^{\Lambda}$ and
$g_\sigma^{\Lambda_c^+} = g_\sigma^{\Lambda}$. The numerical values of the meson-nucleon 
couplings and meson and nucleon masses are from Ref.~\cite{Lalazissis:1996rd} and for the 
meson-$\Lambda$ couplings from Ref.~\cite{Rufa:1990zm} {\textemdash} they are collected in the 
Appendix~B of Ref.~\cite{Fontoura:2015kta}. Presently, not much is known about the effect of SU(4) flavor
symmetry breaking on these effective couplings; recent studies of related couplings (e.g. $g_{N\Lambda K}$ 
and $g_{N\Lambda_c D}$) revealed~\cite{{Khodjamirian:2011sp},{Fontoura:2017ujf}} that the breaking is not 
very large, but a separate study is required to access the effect on the couplings $g_\omega^{\Lambda_c^+}$
and $g_\sigma^{\Lambda_c^+}$. 

The general form of the single-particle wave functions is given in~\ref{A}, where we also present
the values of the corresponding energy eigenvalues. In that same appendix we also
collect the relevant formulae associated with the continuum wave functions for
the ejected proton and $\Lambda_c$. 


\section{Relativistic two-body transition amplitude}
\label{two-body}

\begin{figure}[t]
\begin{center}
\begin{tabular}{cc}
\resizebox{4.5cm}{!}{\includegraphics{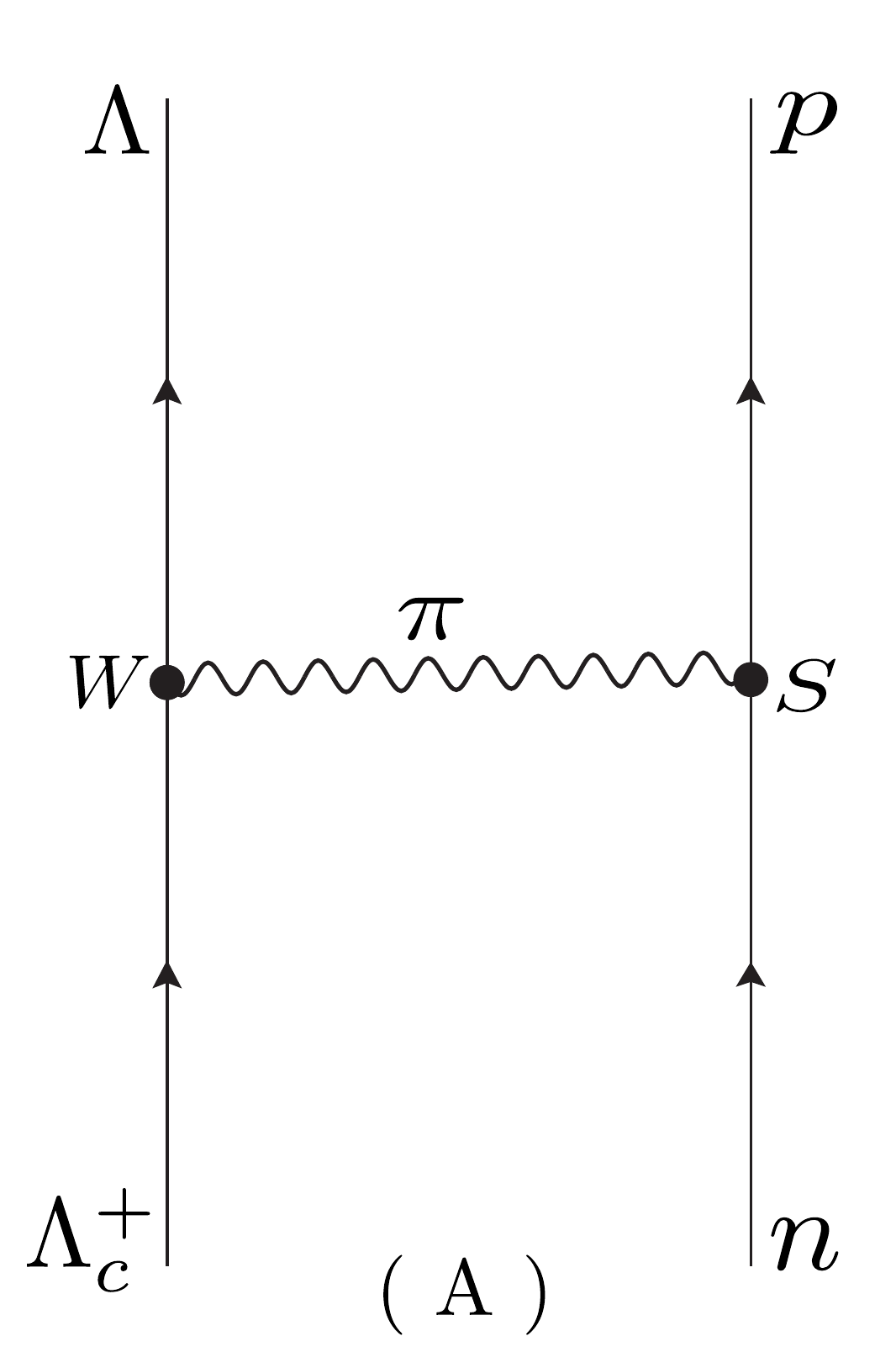}}
&\hspace{2.0cm}\resizebox{4.5cm}{!}{\includegraphics{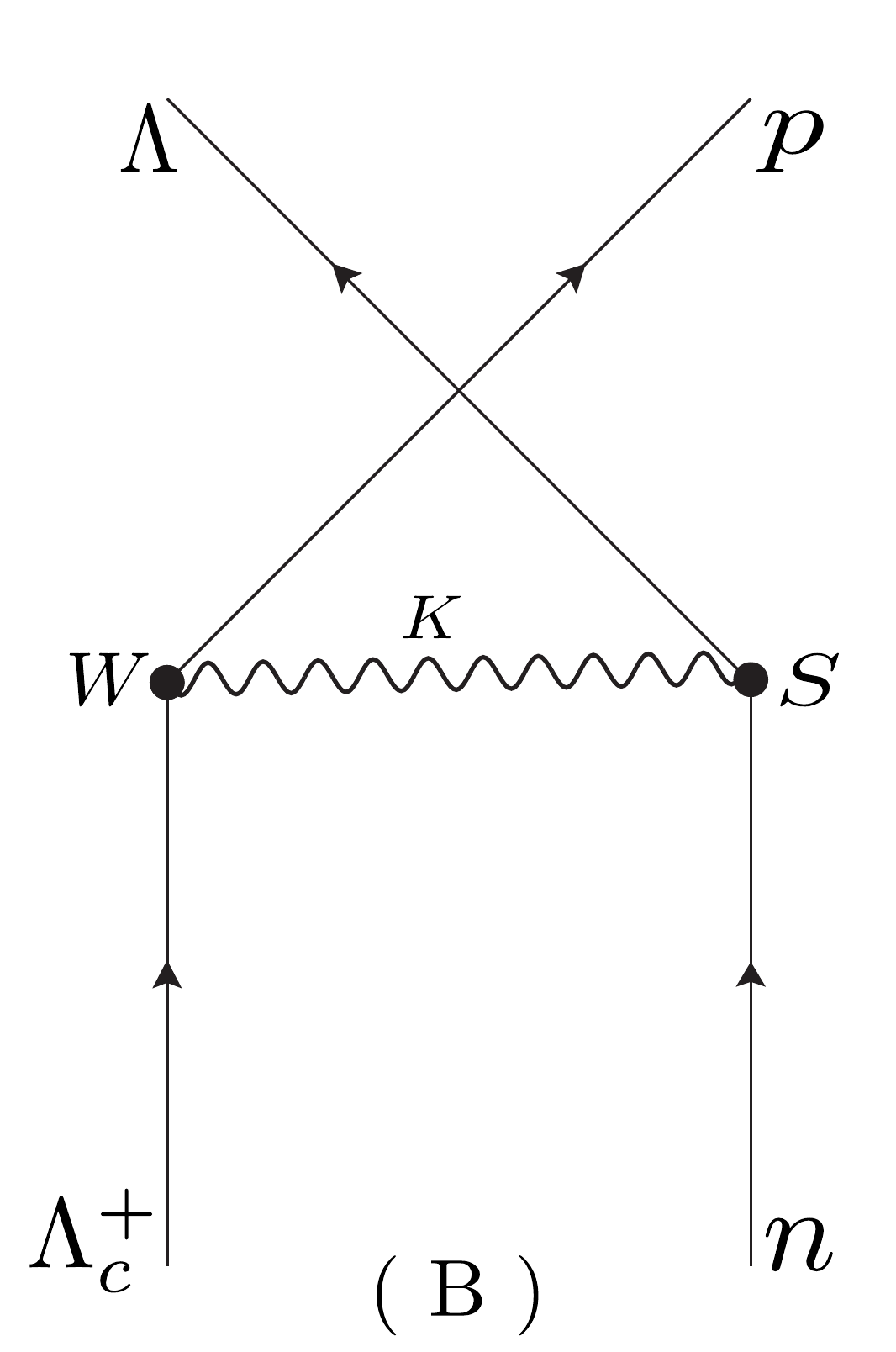}}
\end{tabular}
\caption{One meson exchange diagrams for tree-level processes involving
$\Lambda^{+}_{c}n$ interaction. The diagram $(A)$ corresponds to the one-pion
exchange contribution and the diagram $(B)$ corresponds to the one-kaon exchange
contribution for the two-body transition amplitude.
\label{F1}}
\end{center}
\end{figure}

For the NMWD dynamics we adopt the one-meson-exchange model (OME) including only the pion ($\pi$) 
and kaon ($K$) contributions. Therefore, the transition amplitude $\M$ for the two-body NMWD reaction 
in Eq.~\rf{4} can be obtained from the Feynman rules applied to the two diagrams in Fig.~\ref{F1}.
Due to strangeness and charm selection rules, the $\pi$ meson contributes only 
to diagram (A) and the $K$ meson contributes only to diagram (B).  
The baryon-baryon-meson weak ($W$) and strong ($S$) vertices for $\pi$ and $K$ are taken from the 
corresponding coupling-Hamiltonians, which are:
\bnu
\item 
For one-pion-exchange:
\bea
\mathcal{H}^{S}_{N
N\pi}&=&ig_{NN\pi}\bar{\Psi}_N\gamma_{5}\bm{\tau}\cdot\bm{\Phi}_{\pi}\Psi_{N},
\nn\\
\mathcal{H}^{W}_{\Lambda_{c}\Lambda\pi}&=&
iG_{F}\,m_\pi^{2}\bar{\Psi}_{\Lambda}\left(\,A_\pi+B_\pi
\gamma_{5}\,\right)\bm{\tau}\cdot\bm{\Phi}_{\pi}\Psi_{\Lambda_{c}},
\label{23}
\eea
where $\bm{\tau}$ is the isospin operator, $\bm{\Phi}_{\pi}$ and $\Psi_{N}$ are
pion and nucleon fields, and $A_\pi$ and $B_\pi$ are parity-conserving (PC)
and parity-violating (PV) amplitudes. The strange and charmed baryon fields are written in 
accordance with the isospurion strategy to enforce charge conservation in the 
vertices of Fig.~\ref{F1}, that is,
\beq
\Psi_{\Lambda}=\psi_{\Lambda}\bin{0}{1},
\hspace{0.1cm}
\Psi_{\Lambda_{c}}=\psi_{\Lambda^{+}_{c}}\bin{1}{0}.
\label{24}
\eeq
The strong coupling constant is $g_{NN\pi}=13.3$ and the Fermi coupling constant
is given by $G_F m_\pi^2=2.21\times10^{-7}$. We use the experimental values from
the CLEO collaboration~\cite{Bis95} for the PC and PV amplitudes: 
$A_\pi=-1.56$ and $B_\pi=6.63$ {\textemdash} note that these values are in units of 
$G_F m^2_\pi$, while the CLEO collaboration quotes the results is units of 
$G_F V_{cs} V_{ud} \times 10^{-2}~{\rm GeV}^2$, where $V_{cs}$ and $V_{ud}$ are 
the standard Cabbibo-Maskawa-Kobayashi matrix elements. 

\item 
For one-kaon-exchange:
\bea
\mathcal{H}^{S}_{N \Lambda K}&=&ig_{N\Lambda K}\,\bar{\psi}_{\Lambda}\gamma_{5}\
(\,\Phi^{(K)}\,)^{\dagger}\Psi_{N},
\nn\\
\mathcal{H}^{W}_{\Lambda_cNK}&=&
iG_{F}\,m_\pi^{2}\bar{\psi}_{p}\left(\,A_K+B_K \gamma_5
\right)\phi^{(K^{0})}\psi_{\Lambda^{+}_{c}},
\label{25}
\eea
where $\psi_{\Lambda^{+}_{c}}$ and $\psi_{p}$ are the charmed and proton fields,
$A_K$ and $B_K$ are the PC and PV amplitudes, respectively. For
kaons we can write the field operator $\Phi^{(K)}$ and its hermitian conjugate
$(\,\Phi^{(K)}\,)^{\dagger}$ as
\bea
\Phi^{(K)} &=&\bin{\phi^{(K^{+})}}{\phi^{(K^{0})}},
\hspace{0.2cm}
(\,\Phi^{(K)}\,)^{\dagger} =\cbin{\phi^{(K^{+})^{\dagger}}}
{\phi^{(K^{0})^{\dagger}}}.
\label{26}
\eea
The strong coupling-constant is $g_{N \Lambda K}=-14.1$. 
There are no experimental values for the PV and PC amplitudes $A_K$
and $B_K$; we use the values from theoretical predictions in 
Ref.~\cite{CHENG_TSENG:PRD92} for the $\Lambda^{+}_{c}\rightarrow p+ \bar{K}^{0}$ 
weak transition, namely $A_{K} = - 0.95$ and $B_{K} = 9.17$, in units of $G_F m^2_\pi$. 
\enu

When applying the Feynman rules, the baryon field operators should be expanded
in terms of the eigenfunctions of the corresponding single-particle 
Hamiltonians in Eq.~\rf{9}. Doing this, one gets for the two-body transition amplitude 
\beq
\M(\pb_{\Lambda}s_{\Lambda},\pb_ps_p;j_{\Lambda^{+}_{c}}m_{\Lambda^{+}_{c}},j_nm_n)= 
\M^{\pi} - \M^{K},
\label{27}
\eeq
with
\bea
\hspace{-2.1cm}
\M^{\pi}&=&\int d\xb\,d\yb\,
\bar{\psi}_{\mathbf{p}_{\Lambda}s_{\Lambda}}(\mathbf{x})
\Gamma^{\pi}(t_\Lambda,t_p)
\Psi_{j_{\Lambda^{+}_{c}}m_{\Lambda^{+}_{c}}}(\mathbf{x})
\Delta^{\pi}(|\mathbf{x}-\mathbf{y}|)
\bar{\psi}_{\mathbf{p}_{p}s_{p}}(\mathbf{y})\gamma_{5}\Psi_{j_{n}m_{n}}(\mathbf{y}),
\nn\\
\hspace{-2.1cm}
\M^{K}&=&\int d\xb\,d\yb\,
\bar{\psi}_{\mathbf{p}_{p}s_{p}}(\mathbf{x})
\Gamma^{K}(t_p,t_\Lambda)
\Psi_{j_{\Lambda^{+}_{c}}m_{\Lambda^{+}_{c}}}(\mathbf{x})
\Delta^{K}(|\mathbf{x}-\mathbf{y}|)
\bar{\psi}_{\mathbf{p}_\Lambda s_\Lambda}(\mathbf{y})\gamma_{5}\Psi_{j_{n}m_{n}}(\mathbf{y}),
\label{28}
\eea
where the negative sign in Eq.~\rf{27} comes from the crossing of two fermion  lines  
in  Fig.~\ref{F1}(B). The baryon bound and free Dirac wave-functions 
$\Psi$ and $\psi$ , respectively, have the forms given in Eqs.~\rf{11} and \rf{16}--\rf{18}, 
and we have defined matrices
\bea
\Gamma^{\pi}(t_\Lambda,t_p) &=& \A^{\pi}(t_\Lambda,t_p) + \B^{\pi}(t_\Lambda,t_p) \gamma_5,
\nn\\
\Gamma^{K}(t_p,t_\Lambda) &=& \A^{K}(t_p,t_\Lambda) + \B^{K}(t_p,t_\Lambda) \gamma_5,
\label{29}
\eea
with the pion and kaon effective PC and PV coupling-constants given by
\br
\hspace{-2.0cm}
\A^{\pi}(t_\Lambda,t_p) &=& G_Fm_\pi^2g_{\pi NN}  A_{\pi}  I(t_\Lambda,t_p),
\hspace{0.2cm}
\B^{\pi}(t_\Lambda,t_p) = G_Fm_\pi^2 g_{\pi NN}  B_{\pi}  I(t_\Lambda,t_p),
\nn\\[.1cm]
\hspace{-2.0cm}
\A^{K}(t_p,t_\Lambda) &=& G_Fm_\pi^2 g_{K\Lambda N} A_KK(t_p,t_\Lambda),  
\hspace{.2cm}
\B^{K }(t_p,t_\Lambda) = G_Fm_\pi^2 g_{K\Lambda N} B_KK(t_p,t_\Lambda),
\label{30}
\er
where $I(t_\Lambda,t_p) = 2$ and $K(t_p,t_\Lambda)=1$ are isospin factors.

For the meson propagators, we attach at each vertex the form factor
\brn
F_M(q^2) = \frac{\Lambda_M^2 - m_M^2}{\Lambda_M^2 + q^2},
\ern
with $q^2\equiv (q^0)^2-\bm{q}^2$, where $q$ is the 
transferred momentum, getting
\be
\sDelta^{M}(|\mathbf{x}-\mathbf{y}|)=\int\,\frac{d\qb}{(2\pi)^{3}}
\,\frac{e^{-i\mathbf{q}\cdot(\mathbf{x}-\mathbf{y})}}
{(q^{0})^{2}-\mathbf{q}^{2}-m_{M}^{2}+i\varepsilon}\,
F_M^{2}\left(\,(q^{0})^{2}-\mathbf{q}^{2}\,\right),
\label{31}
\ee
for $M=\pi,K$. These propagators depend on the energy $q^0$ carried by the exchanged meson, 
which is taken as $q^0 = (q_W^0+q_S^0)/2$, with $q_W^0$ and $q_S^0$ fixed by energy 
conservation at the weak ($W$) and strong ($S$) vertices. The numerical values for the cutoffs 
are the same as those used in Ref.~\cite{Fontoura:2015kta}, namely,  
$\Lambda_\pi= 1.3$~GeV and $\Lambda_K=1.2$~GeV.

To conclude this section, we mention that the angular integrations in the transition amplitude 
in Eq.~\rf{27} can be simplified by performing partial-wave expansions. This is done in \ref{B}.

\section{Decay rate}
\label{rate} 

The NMWD rate of a single-$\Lambda^{+}_{c}$ charmed nucleus of baryon number $A$ 
in its ground state with spin $J_I$ and spin-projection $M_I$ and energy $E_{I}$, 
\textit{i.e.}, the partial width for its decay through the reaction in Eq.~\rf{4} into a 
residual nucleus with $(A-2)$ nucleons, emitting a $\Lambda$-hyperon and a proton, 
is given by the Fermi Golden Rule as
\begin{eqnarray}
\Gamma_{nm} &=&\frac{2\pi}{\hat{J}^{2}_I} 
\sum_{M_IJ_FM_F \atop s_\Lambda s_pj_n} 
\int 
\left|\M_A(\pb_\Lambda{s_\Lambda},\pb_p{s_p},j_nJ_FM_F;J_IM_I)\right|^2
\nn\\
&\x&\delta(E_I-E_F-T_R-E_\Lambda-E_p)
\frac{d\pb_\Lambda}{(2\pi)^3}\frac{d\pb_p}{(2\pi)^3}, 
\label{GammaA}
\end{eqnarray}
%
where $\M_A$ is the relativistic nuclear transition amplitude that is specified below, 
$E_F$ and $j_nJ_FM_F$ ($j\equiv (nlj)$) are the energy and quantum numbers of the final 
states of the residual nucleus (\textit{cf.} Eq.~\rf{7}). In addition, ($E_p,\pb_p,s_p$) 
and ($E_\Lambda,\pb_\Lambda,s_\Lambda$) are energies, momenta and spin projections of 
outgoing $p$ and $\Lambda$. There is no summation over isospin projections in Eq.~(\ref{GammaA})
since they have fixed values in the NMWD process in Eq.~\rf{4}, namely $t_p=1/2$ and 
$t_\Lambda = -1/2$. We have included in the energy conservation condition 
the recoil energy
\beq
T_{R}\equiv
T_R(\cos\theta_{\Lambda p})=
\sqrt{M_R^2+p_\Lambda^2+p_p^2+2p_\Lambda\,p_p\cos\theta_{\Lambda p}} - M_R
\label{TR},
\eeq
where $M_R\approx(A-2)M_N$ is the mass of the residual nucleus and 
$\theta_{\Lambda p}$ is the angle between the two outgoing particles.

Within the IPSM, the relations in Eqs.~\rf{5} and \rf{7} allow us to write
\br
E_I &= &E_C + \varepsilon_{j_{\Lambda^{+}_{c}}} + M_{\Lambda^{+}_{c}}, 
\nn\\
E_F &= &E_C - \varepsilon_{j_{n}} - M_{N}, 
\label{49}
\er
where $E_C$ is the energy of the core. Thus,   
the argument of the energy-conserving delta-function in Eq.~\rf{GammaA} reads
\beq
E_I-E_F-T_R-E_\Lambda-E_p=\Delta_{j_{n}}-T_\Lambda-T_p-T_R,
\label{50}
\eeq
where 
\beq
T_i = E_i - M_i \quad\quad (i=\Lambda,p),
\label{51}
\eeq
are kinetic energies, and
\beq
\Delta_{j_{n}}=\Delta_c + \varepsilon_{j_{\Lambda^{+}_{c}}} + \varepsilon_{j_{n}},
\label{52}
\eeq
is the liberated energy. Moreover, from 
\beq
p_{i} = \sqrt{T_{i}\,(T_{i}+2M_{i})\,} \quad\quad(\,i=\Lambda,p\,),
\label{53}
\eeq
one gets
\beq
d{\bf p}_i={p}_i^2\,dp_i\,d\ph_{i} = ( M_{i}+T_i )\,\sqrt{ T_i\,(
2M_{i}+T_i)}\,dT_i\,d\ph_{i}, \quad\quad (i=\Lambda,p),
\label{54}
\eeq
which gives
\br
\Gamma_{nm} = 2\pi\,\int\,\frac{dT_{\Lambda}}{(2\pi)^3}\,\frac{dT_p}{(2\pi)^3}\,
\rho(T_{\Lambda},T_p)
\mathcal{I}(p_{\Lambda},p_{p}),
\label{55}
\er
where 
\beq
\rho(T_{\Lambda},T_p) = (\Mass_{\Lambda}+T_{\Lambda})\,\sqrt{T_{\Lambda}\,
(2\,\Mass_{\Lambda}+T_{\Lambda})}\,(\Mass_p+T_p)\sqrt{T_p\,(2\,\Mass_p+T_p)},
\label{56}
\eeq
and
\bea
\mathcal{I}(p_{\Lambda},p_p)&=&\hat{J}^{-2}_I
\sum_{M_IJ_FM_F \atop s_\Lambda s_pj_n} 
\int d\ph_{{\Lambda}}\,d\ph_{p}\,
\left|\M_A(\pb_\Lambda{s_\Lambda},\pb_p{s_p},j_nJ_FM_F;J_IM_I)\right|^2
\nn\\&\x&
\delta(\Delta_{j_{n}}-T_{\Lambda}-T_p-T_R).
\label{57}
\eea

To conduct the discussion as simply as possible, we will start with 
charmed hypernuclei having a doubly-closed-subshell core (DCSC).
In this case, $J_C=0$, and from Eq.~\rf{7}, the IPSM yields $J_I=j_{\Lambda^{+}_{c}}$, 
$M_I=m_{\Lambda^{+}_{c}}$, $J_F=j_{n}$, and $M_F=m_{n}$. Furthermore, noticing that, 
in the IPSM, such a core functions as the vacuum, $\ket{\underline{0}}$, for particles, 
anti-particles and holes, Eqs.~(\ref{5})-(\ref{7}) take the form
\bea
\ket{J_C} &=& \ket{\underline{0}},
\label{CIF0a}\\
\ket{J_I} &=& 
a^\dagger_{j_{\Lambda^{+}_{c}}}\ket{\underline{0}},
\label{CIF0b}\\
\ket{J_F} &=& \tilde{a}_{j_n}\ket{\underline{0}},
\label{CIF0}
\eea
which imply that the DCSC nuclear transition amplitude $\M_A$ is, 
except for an irrelevant phase-factor, just the two-body transition amplitude 
$\M$ described in Section~\ref{two-body}. Therefore, Eq.~\rf{57} gives
\bea
\hspace{-1.0cm}\mathcal{I}^\mathrm{DCSC}(p_{\Lambda},p_p)&=&
\hat{j}_{\Lambda^{+}_{c}}^{-2}
\sum_{m_{\Lambda^{+}_{c}}j_{n}m_{n}}\sum_{{s_{\Lambda}}{s_p}}
\int d\ph_{{\Lambda}}d\ph_{p}\,
|\M(\pb_{\Lambda}s_{\Lambda},\pb_ps_p;j_{\Lambda^{+}_{c}}m_{\Lambda^{+}_{c}},j_nm_n)|^2
\nn\\
&\x&
\delta(\Delta_{j_{n}}-T_{\Lambda}-T_p-T_R).
\label{59}
\eea

When nuclear recoil is neglected, {\it i.e.} setting $T_R=0$,  we can  use the
completeness relation in Eq.~\rf{21} to integrate over angles in Eq.~\rf{59},
getting
\be
\mathcal{I}^\mathrm{DCSC}(p_{\Lambda},p_p)=
\frac{(4\pi)^4}{\hat{j}_{\Lambda^{+}_{c}}^{2}}
\sum_{\kappa_\Lambda\kappa_pj_nJ}(2J+1)
\left|{\sf M}_J\right|^2
\delta(\Delta_{j_{n}}-T_{\Lambda}-T_p)
\qquad \mbox{(no recoil)},
\label{61} 
\ee
where
\beq
{\sf M}_J={\sf M}^\pi_J-(-)^{j_\Lambda+ j_p+J}{\sf M}^K_J,
\label{62} 
\eeq
is the total angular-momentum-coupled matrix element, the definition and meaning of which 
were explained in \ref{B}. In obtaining this result, we have eliminated the Clebsh-Gordan 
coefficients that appear in Eq.~\rf{36} by performing summations on angular 
momentum projections. 

Thus, from Eq.~\rf{55}, after integrating on $T_p$, we get that, for a DCSC  charmed hypernucleus, 
when described by the  IPSM,  the NMWD rate reads
\be
\Gamma_{nm}^\mathrm{DCSC} = \frac{8}{\pi}
\sum_{\kappa_\Lambda\kappa_pj_nJ}
\frac{\hat{J}^{2}}{\hat{j}_{\Lambda^{+}_{c}}^{2}}
\int_0^{\Delta_{j_n}}dT_\Lambda
\left[\rho(T_{\Lambda},T_p) \left|{\sf M}_J\right|^2\right]_{T_p=\Delta_{j_n}-T_{\Lambda}}
\qquad \mbox{(no recoil)},
\label{63}
\ee
with ${\sf M}_J$ given by Eq.~\rf{62} (for the $\pi+K$ OME model).

From previous works of Refs.~\cite{Kr14,Kr14a,{Fontoura:2015kta},Kr10,Ba08,Go11}, we know that 
to describe the NMWD in $\Lambda$ hypernuclei with open-shell cores within the IPSM it is enough 
to make the  replacement $\hat{J}^{2}/\hat{j}^{2}_{\Lambda} \go F^{j_N}_J$ in the decay-rate expression 
with the DCSC, where $F^{j_N}_J$ is a spectroscopic factor. Making the same replacements here, {\it i.e.}
$\hat{J}^{2}/\hat{j}^{2}_{\Lambda^{+}_{c}} \go F^{j_N}_J$ in Eq.~\rf{63}, with  $F^{j_N}_J$
given by Eq.~\rf{8}, we get  that NMWD rate in recoilless charmed hypernuclei 
is given by
\bea
\hspace{-1.0cm}
\Gamma_{nm} &=& \frac{8}{\pi } \sum_{\kappa_\Lambda\kappa_pj_nJ}
F^{j_n}_J
\int_0^{\Delta_{j_n}}dT_{\Lambda}dT_p \, \rho(T_{\Lambda},T_p) 
\left|{\sf M}_J\right|^2\delta(\Delta_{j_{n}}-T_{\Lambda}-T_p),
\nn\\
&=& \frac{8}{\pi }
\sum_{\kappa_\Lambda\kappa_pj_nJ}
F^{j_n}_J
\int_0^{\Delta_{j_n}}dT_{\Lambda}
\left[\rho(T_{\Lambda},T_p) \left|{\sf M}_J\right|^2\right]_{T_{p}=\Delta_{j_n}-T_\Lambda}
\hspace{0.2cm}\mbox{(no recoil)},
\label{64}
\eea
for both  open- and closed-subshell cores.

\section{Effect of nuclear recoil}
\label{recoil}

When including recoil, one needs to perform the angular integration in Eq.~\rf{57}.
We proceed as indicated in Eqs.~(55)-(57) of Ref.~\cite{Fontoura:2015kta}, in that one
makes the replacement
\be
\int dT_{\Lambda}dT_p 
\delta(\Delta_{j_{n}}-T_{\Lambda}-T_p)
\go \fot\int d\cos\theta_{\Lambda p}dT_{\Lambda}dT_p \, 
\delta(\Delta_{j_{n}}-T_{\Lambda}-T_p-T_R) \, \cdots ,
\label{65}
\ee
in the first branch of Eq.~\rf{64}, getting
\be
\Gamma_{nm} = \frac{4}{\pi }
\sum_{\kappa_{\Lambda}\kappa_{p\,}j_nJ}
F^{j_n}_J\,\int\,d\cos\theta_{\Lambda p}\,dT_{\Lambda}\,dT_{p}\,
\rho(T_{\Lambda},T_{p})\,\left|{\sf M}_J\right|^2\,\delta[f(T_{p})],
\label{66a}
\ee
where $f(T_{p})$ is given by
\be
f(T_{p}) = \Delta_{j_{n}}-T_{\Lambda} - T_p - T_R .
\label{f}
\ee

Details on the evaluation of the integration over $T_p$ are presented in~\ref{C}. 
The final result for the rate can be expressed as 
\bea
\Gamma_{nm} &=&
\frac{4}{\pi}\sum_{\kappa_{\Lambda}\kappa_{p}j_{n}J}\,
F^{j_n}_J\,
\int dT_\Lambda\,
\int d\cos\theta_{\Lambda p}\,
\theta(K_2)
\nonumber\\&\times&
\left\{
\left[\,\frac{\rho(T_{\Lambda},T_p)}{|f'(T_p)|}
\theta(T_p)\, \theta_0[f(T_p)]
\,\left|{\sf M}_J\right|^2
\right]_{T_p\to T_p^+}
+
\left[\;\cdot\;\right]_{T_p\to T_p^-}
\right\},
\label{rate3}
\eea
where $T^{\pm}_p$ is given by
\bea
T_{p}^\pm &=& 
\frac{K_1(T_\Lambda,\cos\theta_{\Lambda p}) \pm 
|p_\Lambda\cos\theta_{\Lambda p}| \sqrt{K_2(T_\Lambda,\cos\theta_{\Lambda p}})}
{2K_3(T_\Lambda,\cos\theta_{\Lambda p})},
\label{rootsNMWD}
\eea
where $K_1$, $K_2$ and $K_3$ are given by Eqs.~(\ref{K1})-(\ref{K3}).
Here, the step functions ensure positivity of $K_2$ and $T_p$, and 
$\theta_0(x)$  
\be
\theta_0(x) = 
\left\{
\begin{array}{ll}
1 & \mbox{if}\; |x| < \epsilon
\\
0 & \mbox{otherwise},
\end{array}
\right.
\label{theta0}
\ee
with $\epsilon$ being a suitably chosen small positive value, ensures that 
the roots are not spurious solutions to $f(T_{p})=0$ {\textemdash} see \ref{C}
for details. The intervals of integration in Eq.~\rf{rate3} are 
$-1 < \cos\theta_{\Lambda p} < 1$ and $0 < T_\Lambda < T_\Lambda^\mathrm{max}$ 
with 
\be
T_\Lambda^\mathrm{max} = 
\frac{(2M_p+\Delta_{j_n})\Delta_{j_n}}{2(M_p+M_\Lambda+\Delta_{j_n})}
\approx \frac{(2M_p+\Delta_c)\Delta_c}{2(M_p+M_\Lambda+\Delta_c)} 
= 553\,\text{MeV}.
\label{TLambdamax}
\ee

In what follows, we shall make reference to the $\Lambda$ kinetic energy spectrum 
and to the pair opening-angle distribution, which are, respectively, the partial integrands 
in the variables $T_\Lambda$ and $\cos\theta_{\Lambda p}$ and are denoted by
\be
S(T_\Lambda)=\frac{d\Gamma_{nm} }{dT_\Lambda},
\label{81}
\ee 
and 
\be
S(\cos\theta_{\Lambda p})=\frac{d\Gamma_{nm} }{d\cos\theta_{\Lambda p}}.
\label{82}
\ee 

%
\section{Numerical Results and Discussion}
\label{results}

We start presenting numerical results for the rate of the one-neutron-induced NMWD 
of $_{\Lambda^+_c}^{12}$N. We concentrate on this particular nucleus to compare results 
with the study of the $_{\Lambda}^{12}$C hypernucleus we have performed in 
Ref.~\cite{Fontoura:2015kta}. As discussed in the previous sections, we employ the 
IPSM and consider the $\pi$ and $\pi+K$ OME model for the weak decay process.  In this model, 
the neutron states contributing to the transition are the $j_n=1s_{1/2}$ and $1p_{3/2}$, 
and the $\Lambda^+_c$ is always considered to be in the state $j_{\Lambda^+_c}=1s_{1/2}$. 

We remark that we have found numerically that the contribution from second term in Eq.~\rf{rate3}, 
coming from the root $T_p= T^-_p$, is relatively small and may be neglected for all practical 
purposes. This same feature was seen in the nonrelativistic treatment of the recoil effect in the
NMWD of the $_{\Lambda}^{12}$N hypernucleus; specifically, Eq. (63) of Ref.~\cite{Fontoura:2015kta}. 
Here and there, this feature can be attributed to phase-space.

\begin{table}[h]
\caption{Parity-conserving (PC), parity-violating (PV),  and  total decay
rates for the nonmesonic weak decay $^{12}_{\Lambda^{+}_{c}}{\rm N}\go {^{10}{\rm C}}+p+\Lambda$, 
in units of $10^{-11} \text{MeV}$. 
Three approaches to recoil effects and two choices of OME models are considered.}
\begin{center}
\begin{tabular}{c|cccccc}
\hline \\[-0.4cm]
\hspace{.1cm} Model\hspace{.1cm}&\hspace{.1cm}
$\Gamma^{\text{PC}}_{nm}$\hspace{.1cm}&\hspace{.1cm}
$\Gamma^{\text{PV}}_{nm}$\hspace{.1cm}&\hspace{.1cm}
$\Gamma_{nm}$
\\ \hline
\hspace{-2.5cm}
No Recoil\\
\hspace{-2.8cm}
$\pi\hspace{1.cm}$
\hspace{.1cm}&\hspace{.1cm}
$1.31$\hspace{.1cm}&\hspace{.1cm}
$0.89$\hspace{.1cm}&\hspace{.1cm}
$2.20$
\\[-0.1cm] \hspace{-2.8cm}
$\pi+K\hspace{0.1cm}$
\hspace{-.1cm}&\hspace{.1cm}
$1.94$\hspace{.1cm}&\hspace{.1cm}
$0.90$\hspace{.1cm}&\hspace{.1cm}
$2.84$
\\[-0.1cm] \hline
\hspace{-1.cm}
Relativistic Recoil\\
\hspace{-2.7cm}
$\pi\hspace{1.cm}$
\hspace{.1cm} &\hspace{.1cm}
$0.91$\hspace{.1cm}&\hspace{.1cm}
$0.59$\hspace{.1cm}&\hspace{.1cm}
$1.50$\\[-0.1cm]
\hspace{-2.7cm}
$\pi+K\hspace{0.1cm}$
\hspace{.1cm}&\hspace{.1cm}
$1.47$\hspace{.1cm}&\hspace{.1cm}
$0.60$\hspace{.1cm}&\hspace{.1cm}
$2.07$
\\[-0.1cm] \hline
\hspace{-.3cm}
Nonrelativistic Recoil\\
\hspace{-2.8cm}
$\pi\hspace{1.cm}$
\hspace{.1cm}&\hspace{.1cm}
$1.01$\hspace{.1cm}&\hspace{.1cm}
$0.68$\hspace{.1cm}&\hspace{.1cm}
$1.70$
\\[-0.1cm]
\hspace{-2.7cm}
$\pi+K\hspace{0.1cm}$
\hspace{.1cm}&\hspace{.1cm}
$1.55$\hspace{.1cm}&\hspace{.1cm}
$0.69$\hspace{.1cm}&\hspace{.1cm} $2.24$
\\
\hline
\end{tabular}
\end{center}
\label{tab2}
\end{table}

In Table~\ref{tab2} we present the different contributions to the rates. We consider separately 
contributions coming from the parity-conserving ($\Gamma^{\text PC}_{nm}$) and parity-violating 
($\Gamma^{\text PV}_{nm}$) transitions, that correspond, respectively, to the $B$ and $A$ terms
in Eqs.~(\ref{23}) and (\ref{25}), and also give the total rate, $\Gamma_{nm} = \Gamma^{\text PC}_{nm}
+ \Gamma^{\text PV}_{nm}$, all of them are in units of $10^{-11} \text{MeV}$. We recall that 
the PC contribution to $\Gamma_{nm}$ comes from the amplitudes ${\cal A}^\pi$  and ${\cal A}^K$
and the PV contribution comes from ${\cal B}^\pi$ and ${\cal B}^K$, defined in Eq.~(\ref{30}).   
For each of these quantities we present three different sorts of results: first, results for the 
decay rates without recoil effects, computed with Eq.~\rf{64}; second, results with relativistic 
recoil effects, computed with Eq.~(\ref{rate3}); and, finally, results with nonrelativistic recoil 
effects, in which a nonrelativistic approximation is made for the recoil energy $T_R$, following 
the procedure in Ref.~\cite{Fontoura:2015kta} for the $_{\Lambda}^{12}$C hypernucleus. 

The following conclusions can be drawn from the results displayed in the Table:
\begin{enumerate}
\item 
The contribution of the one kaon exchange potential is quite significant for the PC decay rate, but it is 
very small for the PV decay.
\item
Recoil has a sizable impact on the rate and goes in the direction of decreasing it, at the level of 20\% to 
30\%. 
\item
The difference between the results with relativistic and nonrelativistic treatment of recoil effects 
are at the level of 10\%, surprisingly not a large effect.
\item
The predicted NMWD rates $\Gamma_{nm}$ are of the same order of magnitude as the partial decay rate for the 
corresponding mesonic decay $\Gamma(\Lambda^{+}_c\rightarrow\Lambda +\pi) = (4.28 \pm 0.26) \x 10^{-11}$ MeV, 
whose measured branching fraction, calibrated relative to the $pK^-\pi^+$ mode, is $B(\Lambda_c^+ \to 
\Lambda+\pi^+)=(1.07\pm0.28)\%$~\cite{PDG}. In Ref. ~\cite{Bis95}, is it reported the value 
$\Gamma(\Lambda^{+}_c\rightarrow\Lambda +\pi)=(0.40\pm  0.11)\x10^{11} \rm{s}^{-1}= (2.63\pm  0.72)\x10^{-11} \rm{MeV}$. 
\end{enumerate}

\begin{figure}[htb]
\begin{center}
\includegraphics[scale=0.4]{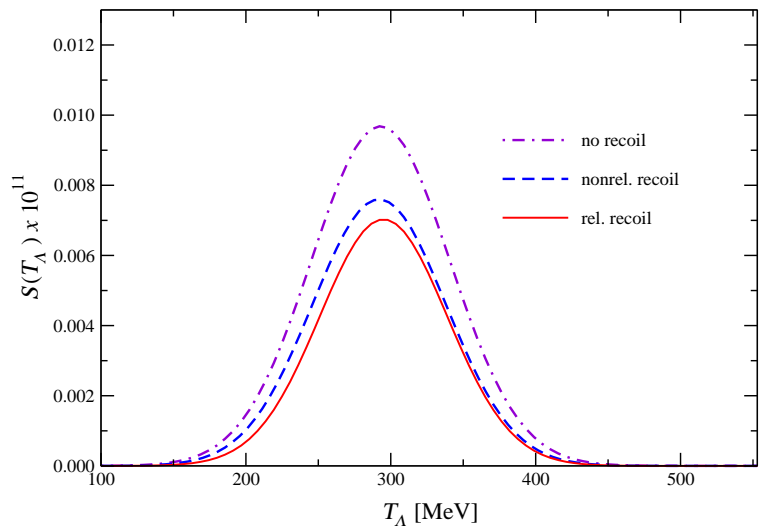}
\end{center}
\vspace{-0.8cm}
\caption{The $_{\Lambda^+_c}^{12}$N NMWD  spectrum as a function of the kinetic energy  $T_\Lambda$, evaluated  
without recoil (dash-dotted violet-line), with nonrelativistic recoil (dashed blue-line) and with relativistic 
recoil (solid red-line).}
\label{fig2}
\end{figure}

In Fig.~\ref{fig2} we show the $\Lambda$ kinetic energy spectrum $S(T_\Lambda)$, defined in Eq.~\rf{81}.
The figure shows the spectra evaluated without recoil, with fully relativistic recoil and with 
nonrelativistic recoil. Independently of how the recoil effect is treated, the kinetic energy spectra 
$S(T_\Lambda)$ always have a symmetric bell shape, with centroid at about $T_\Lambda=300$ MeV, which 
is roughly half the maximum energy $T_\Lambda^\mathrm{max} $ of the emitted $\Lambda$.

\begin{figure}[hbt]
\begin{center}
\includegraphics[scale=0.4]{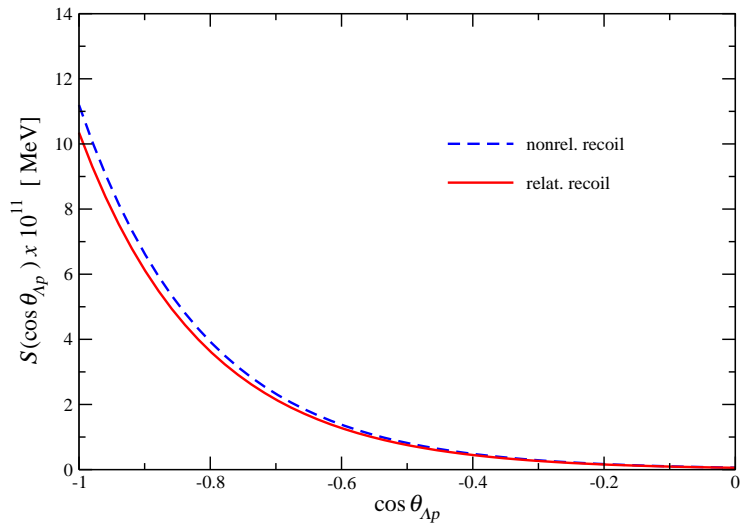}
\end{center}
\vspace{-0.8cm}
\caption{Opening-angle distribution of the emitted $\Lambda$-$p$ pair in the $_{\Lambda^+_c}^{12}$N NMWD, 
evaluated with nonrelativistic recoil (dashed blue-line) and with relativistic recoil (solid red-line).}
\label{fig2a}
\end{figure}

Finally, in Fig.~\ref{fig2a} we show the results for the opening-angle distribution 
$S(\cos\theta_{\Lambda p})$, defined in Eq.~\rf{82}. We recall that the angular distribution of the 
emitted particles is due to recoil; when recoil is neglected, the particles are emitted back to back. 
We show results calculated with relativistic and nonrelativistic expressions for the recoil. The 
opening-angle distribution $S(\cos\theta_{\Lambda p})$ in Fig.~\ref{fig2a} 
is similar to the analogous one $S_p (\cos\theta_{n p})$ of Fig.~\ref{fig2} in Ref.~\cite{Go11}, in that 
it has a maximum for $\cos\theta_{\Lambda p}=-1$, but it extends a little further towards smaller angles.

%
\section{Conclusions and Perspectives}
\label{sec7}

The investigation of the production of heavy flavor hadrons containing a charm-quark and their 
interaction with ordinary hadrons in nuclear medium is of considerable contemporary interest once it provides 
an additional means for a better understanding of new forms of nuclear matter~\cite{{Briceno:2015rlt},{Hosaka:2016ypm}}. 
A major difficulty in such an investigation program is the lack of experimental information on the 
free-space and in-medium interactions involving charmed hadrons unlike what happens in similar problems 
involving strange hadrons. In a situation with a lack of experimental information, one way to proceed in 
model building is to use symmetry constraints and analogies with other similar processes. With such a motivation, 
we investigated in this work the nonmesonic weak decay of nuclei containing a single $\Lambda_{c}^{+}$,  
relying on previous experience with the analogous $\Lambda$ hypernucleus. 

We have discussed in Sections~\ref{IPSM} and \ref{two-body} the extension to the
$^{12}_{\Lambda^{+}_{c}}{\rm N}$ hypernucleus of a relativistic formalism previously developed in 
Ref.~\cite{Fontoura:2015kta} for the NMWD of the $^{12}_{\Lambda}{\rm C}$ hypernucleus. We have worked 
within the framework of the IPSM, with the dynamics of the $\Lambda^{+}_{c}n\to\Lambda p$ decay
described by the $(\pi, K)$ OME model, with unknown couplings fixed by $SU(4)$ flavor symmetry. 
Dirac plane waves were expanded in spherical partial waves, meson propagators were multipole-expanded, 
so that the two-body transition matrix elements of the transition could be expressed in terms of 
two-dimensional radial integrals.
Next, in Section~\ref{rate}, we have implemented the formalism for hypernuclei whose cores have only 
closed subshells. Then, to make contact with previous calculations, we have neglected nuclear recoil 
effects, which allowed us to reduce six-dimensional momentum-space integrals to simple one-dimensional integrals. 
The result obtained within this approximation was then generalized to include hypernuclei with open-shell 
cores. Finally, in Section~\ref{recoil}, we have implemented a fully relativistic treatment of the recoil 
effects. Our results have shown that nuclear recoil has a sizable impact on the decay rate, and goes in the 
direction of decreasing it by 20\% to 30\%. Nuclear recoil gives an angular distribution to the emitted
$\Lambda p$ pair and impacts significantly single kinetic energy spectra. 

Very recently, the authors of Ref.~\cite{Steinheimer:2016jjk} have suggested that small-sized 
(typically with $A=3$) charmed hypernuclei can be formed in nuclear collisions and identified 
through their mesonic decays~\cite{Ghosh:2016jjv} {\textemdash} in this respect, a recent exact 
three-body calculation~\cite{Garcilazo:2015qha} using baryon-baryon interactions obtained from a 
chiral constituent quark model predicts a $J=3/2$ charmed hypertriton with a binding energy between 
$140$ and $715$~MeV. Our study, on the other hand, suggests 
identification of the formation of charm hypernuclei via nonmesonic 
weak decays, which are unique, as they can only occur in the nucleus. Moreover, one of the
the most interesting aspects of our results is that the predicted NMWD rate is of the same order 
of magnitude as the measured  decay rate for the corresponding weak mesonic decay:
\be
\Gamma_{nm}(^{12}_{\Lambda^{+}_{c}}{\rm N}\go {^{10}{\rm C}}+p+\Lambda)
\simeq \Gamma(\Lambda_c^+\to \Lambda+\pi^+).
\label{83}
\ee 
Branching ratios even smaller than $B(\Lambda_c^+\to \Lambda+\pi^+)=(1.07\pm0.28)\%$ have been 
measured~\cite{PDG}. This  suggests  that once the charmed hypernucleus $^{12}_{\Lambda^{+}_{c}}{\rm N}$ 
is produced, its NMWD should be measurable. We have limited our discussion to  $^{12}_{\Lambda^{+}_{c}}{\rm N}$, 
but it is expected that the NMWD will be very similar in other $\Lambda_c$ hypernuclei, since this is the case 
in $\Lambda$ hypernuclei, as can be seen in Table II in Ref.~\cite{Kr10a}. This makes it even more feasible 
to detect the $\Lambda p$ pair in the final state. Needless to say that knowledge of kinetic energy spectra 
and of opening-angle $\Lambda p$  correlations such as  those shown in Figs.~\ref{fig2} and \ref{fig2a}  
would be of help in this search. One should be aware, however, of the difficulties involved in the 
identification of a charmed nucleus like $^{12}_{\Lambda^{+}_{c}}{\rm N}$ in a $\bar{p}-$nucleus 
collision. One possibility would be the production through the reaction chain $\bar p + p \rightarrow D^{+} 
+ D^{-}$, $D^{+} + ^{12}\textrm{C} \rightarrow ^{12}_{\Lambda^+_c}\textrm{N} + \pi^0$. The difficulty here
is related to the detection of the $\pi^0$ in the final state. On the other hand, the direct process
$\bar p + p \rightarrow \Lambda^+_c + \bar{\Lambda}^{-}_c$, as suggested in Ref.~\cite{Shyam:2016uxa},
would produce a proton hole in $^{12}\textrm{C}$, giving rise to the charmed hipernucleus with 
$A=12$ and $Z=6$, i.e. a $^{12}_{\Lambda^{+}_{c}}{\rm C}$ with six neutrons, five protons and one
$\Lambda^+_c$. We reserve for a future publication the investigation of the NMWD of this nucleus.   

To conclude, we mention that our study is a first incursion in the study of NMWD of charmed 
hypernuclei. We have limited the study to a two-body final state in the decay of $\Lambda^+_c$
but, of course, decay processes with larger branching ratios involving multiparticle final states 
should be explored in the identification of the formation and decay of charmed hypernuclei. 
Therefore, one can envisage a long path, both in theory and experiment, in the production
of these fascinating new forms of nuclear matter.

\ack{
Work partially supported by the Brazilian agencies Conselho Nacional de Desenvolvimento Cient\'{\i}fico 
e Tecnol\'ogico - CNPq, Grants 150659/2015-6 (CEF) and 305894/2009-9 (GK) and Funda\c{c}\~ao de Amparo 
\`a Pesquisa do Estado de S\~ao Paulo - FAPESP,  Grant 2013/01907-0 (GK), 
as well as by the Argentinean agencies Consejo Na\-cio\-nal de In\-ves\-ti\-ga\-ciones
Cient\'ificas y T\'ecnicas - CONICET, Grant No. PIP 0377 (FK), and Fondo para la
Investigaci\'on Cient\'ifica y Tecn\'ologica - FONCYT, Grant No. PICT-2010-2680 (FK).
}

\appendix

\section{Bound and continuum single-particle wave functions}
\label{A}

For completeness and to make the paper self-contained, we collect here the relevant formular
associated with the bound and continuum single-particle wave functions. The bound single-particle
wave functions have the general form
\beq
\Psi_{\kappa m}(\rb)
=\frac{1}{r}\bin{F_{\kappa}(r)\Phi_{\kappa m}(\rh)}{-iG_{\kappa}(r)
\Phi_{-\kappa m}(\rh)}\equiv
\bin{\Psip_{\kappa m}(\rb)}{-i\Psin_{\kappa m}(\rb)}.
\label{11}
\eeq
with $\kappa=\pm1,\pm2,\dots$, $j_\kappa=|\kappa|-1/2$,
\beq
l_{\kappa} = \left\{ \begin{array}{ll}\kappa & \mbox{for $\kappa>0$,} \\
-\kappa-1 & \mbox{for $\kappa<0$.} \end{array} \right.
\label{12}
\eeq
The angular part is written, in standard notation, as
\beq
\Phi_{\kappa m}(\rh)=\sum_{s\mu} (l\mu\fot s|jm) \, Y_{l\mu}(\rh)\chi_s.
\label{13}
\eeq
The radial part is determined from the eigenvalue equation
\beq
\hat{h}\, \Psi_{\kappa m}(\rb) = (M + \varepsilon_\kappa)\, \Psi_{\kappa m}(\rb),
\eeq
where $\varepsilon_\kappa$ are the single-particle energies (s.p.e.), and the normalization
is
\beq
\int d\rb\; \Psi_{\kappa m}^\dagger(\rb)\, \Psi_{\kappa m}(\rb) = 1.
\eeq
%

\begin{table}[H]
\caption{Single-particle energies obtained from the MFA for $^{12}{\rm C}$. 
For the experimental values for $^{12}\textrm{C}$, see the explanation in the 
text. The last line gives the ${1s}_{1/2}$ single-particle energy for the 
$\Lambda^+_c$ in $^{12}_{\Lambda^+_c}{\rm N}$ hypernucleus (composed
by 6 protons, 5 neutrons and 1 $\Lambda^+_c$). 
All energies are in MeV.
\label{tab:spe}}
\begin{center}
\begin{tabular}{c|rrr}
\hline
\\[-.4cm]\hspace{.1cm}Energy Level\hspace{.1cm}&\hspace{.3cm}
Calculated\hspace{.1cm}&\hspace{.4cm}Experiment $^{12}\textrm{C}$ 
\\ \hline
\hspace{.3cm} 
${\mbox{1s}_{1/2}}_p$
\hspace{.1cm}&\hspace{-.1cm}
$-38.53$\hspace{.1cm}&\hspace{.1cm}$-33.5$&
\\
\hspace{.3cm} 
${\mbox{1p}_{3/2}}_p$
\hspace{.1cm}&\hspace{.1cm}
$-13.52$\hspace{.1cm}&\hspace{.1cm}$-15.96$&
\\
\hspace{.3cm} 
${\mbox{1s}_{1/2}}_n$
\hspace{.1cm}&\hspace{.1cm}
$-42.03$\hspace{.1cm}&\hspace{.1cm}$-36.3$&
\\[0.1cm]
\hspace{.3cm} 
${\mbox{1p}_{3/2}}_n$
\hspace{.1cm}&\hspace{.1cm}
$-16.65$\hspace{.1cm}&\hspace{.1cm}$-18.72$&
\\[0.1cm]
\hline
\\[-0.4cm]
\hspace{.7cm} 
${\mbox{1s}_{1/2}}_{\Lambda^{+}_{c}}$ \; in \; $^{12}_{\Lambda^+_c}\textrm{N}$
\hspace{.1cm}
&\hspace{.1cm}$-14.32$\hspace{.1cm}&\hspace{.1cm}$-$
\\[0.2cm]
\hline
\end{tabular}
\end{center}
\end{table}

Table~\ref{tab:spe} presents the results for the single-particle energies (s.p.e.) {\textemdash} 
the values of the parameters are fixed as discussed at the end of Sec.~{ISPM}. The experimental 
values for neutron and proton $1p_{3/2}$ s.p.e. in $^{12}\textrm{C}$ are taken to be 
$\mbox{1p}_{3/2}$ 
hole states obtained from the separation energies calculated from the differences of experimental
binding energies of $^{12}\textrm{C}$, $^{11}\textrm{C}$ and $^{12}\textrm{B}$~\cite{exp-B}
\footnote{Strictly speaking, the s.p.e. are equal to the separation energies 
only for states double-closed-shell nuclei that lie close to the Fermi level~\cite{Boh69}.
Therefore, the experimental values quoted in Table~\ref{tab:spe} should be taken as guidance 
only.}: $\mathcal{B}(^{12}{\rm C})=92.16279$~MeV, $\mathcal{B}(^{11}{\rm C})=73.4414$~MeV, 
and $\mathcal{B}(^{11}{\rm B})=76.2059$~MeV. The proton $\mbox{1s}_{1/2}$ s.p.e. 
energy is obtained from data~\cite{Yo03} on the knock-out reaction $^{12}$C(p,2p)$^{11}$B which 
indicate that the deep-proton-hole state $\mbox{1s}_{1/2}$ is located $\sim 20$ MeV above the 
$\mbox{1p}_{3/2}$ ground state in $^{11}$B, giving a value of $\sim -33.5$ MeV for 
${\mbox{1s}_{1/2}}_p$ in $^{12}{\rm C}$. However, knock-out reactions on neutrons such as (p,p'n) 
and (e,e'n) have not been reported, therefore, we assume that the energy separation between neutron 
$\mbox{1p}_{3/2}$ and $\mbox{1s}_{1/2}$ states is the same as that of the protons, which yields the 
value of $\sim -36.3$ MeV for the the ${\mbox{1s}_{1/2}}_n$ state (note that in our calculation of 
the s.p.e., the effect of the Coulomb force). 

As remarked in Sec.~\ref{IPSM}, to describe the NMWD of $^{12}_{\Lambda_{c}^+}{\rm N}$ in the 
IPSM, the MFA  is performed for $^{12}{\rm C}$, i.e. the $\Lambda^+_c$ wave functions should be
those generated by mean fields for the $^{12}{\rm C}$ nucleus. However, it is instructive to 
estimate the effect of neglecting the contribution of $\Lambda^{+}_{c}$ to the 
sources of the meson mean fields. We have repeated the calculation of the bound wave functions 
by solving the Dirac and meson equations self-conistently, but still enforcing spherical symmetry
of the nucleus. We found that the biding energies of the ${\mbox{1s}_{1/2}}_p$ and 
${\mbox{1p}_{3/2}}_p$ states decrease by 10\% and 20\% respetively, of the $\Lambda^+_c$ is 
increased by 6\%, and there is almost no change in the neutron binding energies. On the other 
hand, the changes in the bound-state wave functions lead o a change of the order of 1\% in NMWD 
rates. Deviations from non-sphericity of the nuclei have not been estimated and their study
is left for a future publication.

The continuum single-particle states should be taken as the positive-energy 
scattering eigenfunctions of the Hamiltonian in Eq.~\rf{9} with asymptotic momentum 
$\pb$ and spin projection~$s$. However, those will be approximated by the corresponding Dirac 
plane waves, which are expanded as follows~(see Ref.~\cite{Do85}, Appendix D):
\br
\psi_{\pb s}(\rb) 
= \sum_{\kappa m} \ov{\ph s}{\kappa m}^\ast \,
\psi_{p\kappa m}(\rb),
\label{16}
\er
with
\beq
\ov{\ph s}{\kappa m}^\ast
=4\pi i^l\sum_{\mu}(l\mu\fot s|jm)Y^*_{l\mu}(\ph),
\label{17}
\eeq
and
\beq
\psi_{p\kappa m}(\rb) =  
\bin{f_{p\kappa}(r)\Phi_{\kappa m}(\rh)}{-ig_{p\kappa}(r)\Phi_{-\kappa m}(\rh)}
\equiv
\bin{\psip_{p\kappa m}(\rb)}{-i\,\psin_{p\kappa m}(\rb)},
\label{18}
\eeq
where the radial partial-waves, in unitary normalization, are
\beq
f_{p\kappa}(r)=\sqrt{\frac{E+\Mass_N}{2E}}j_{l_\kappa}(pr),
\label{19}
\eeq
and
\beq
g_{p\kappa}(r)=
-\sgn(\kappa)\sqrt{\frac{E-\Mass_N}{2E}}j_{{\bar l}_\kappa}(pr),
\label{20}
\eeq
with ${\bar l}_\kappa=l_{-\kappa}$. 
The expansion coefficients $\ov{\ph s}{\kappa m}^\ast $ fulfill the following 
relations
\be
\sum_s\int d\ph\,
\ov{\ph s}{\kappa m}^\ast \ov{\ph s}{\kappa' m'}
=(4\pi)^2\delta_{\kappa\kappa'}\delta_{mm'},
\label{21}
\ee
and
\be
2\hat{j}^{2}\delta_{jj'}\hspace{-.1cm}\sum_{s m}\int d\ph\,
\ov{\ph s}{\kappa m}^\ast \ov{\ph s}{\kappa' m} \cdots =
(4\pi)^2\delta_{\kappa\kappa'}\int_{-1}^1 d\cos\theta \cdots\quad\quad.
\label{22}
\ee
The first of these relations can be easily verified, while the second one is 
shown in~Appendix~A of Ref.~\cite{Fontoura:2015kta}.

\section{Partial-wave decomposition of ${\cal M}^\pi$ and ${\cal M}^K$}
\label{B}

 Using Eq.~\rf{16} for both outgoing particles, the transition 
amplitude in Eq.~\rf{27} becomes
\bea
\M = \underset{\kappa_p m_p}{\underset{\kappa_\Lambda m_\Lambda}{\sum}}
\ov{\ph_\Lambda s_\Lambda}{\kappa_\Lambda m_\Lambda} \, \ov{\ph_p s_p}{\kappa_p m_p} 
({\sf M}^\pi-{\sf M}^K),
\label{34}
\eea
where 
\br
\hspace{-2.5cm}
{\sf M}^\pi
&\equiv&\int d\xb d\yb \bar{\psi}_{p_{\Lambda}\kappa_{\Lambda} 
m_{\Lambda}}(\xb)\Gamma^\pi(t_\Lambda,t_p){\Psi}_{j_{\Lambda^{+}_{c}} 
m_{\Lambda^{+}_{c}}}(\xb){ \sDelta}^\pi(|\xb-\yb|)  \bar{\psi}_{p_p\kappa_p m_p}(\yb) 
\gamma_{5} {\Psi}_{j_nm_n}(\yb),
\nn\\
\hspace{-2.5cm}
{\sf M}^{K}&\equiv&\int d\xb d\yb
\bar{\psi}_{p_p\kappa_p m_p}(\mathbf{x})
\Gamma^{K}(t_p,t_\Lambda)
\Psi_{j_{\Lambda^{+}_{c}}m_{\Lambda^{+}_{c}}}(\mathbf{x})\sDelta^{K}(|\mathbf{x}-\mathbf{y}|)
\bar{\psi}_{p_{\Lambda}\kappa_{\Lambda} m_{\Lambda}}(\mathbf{y})
\gamma_{5}\Psi_{j_{n}m_{n}}(\mathbf{y}).
\label{35}
\er
Now we introduce the angular momentum couplings ${\bm J}= {\bm j}_{\Lambda^{+}_{c}} + {\bm j}_n$ and
${\bm J}'= {\bm j}_\Lambda + {\bm j}_p$ in ${\sf M}^\pi$, and ${\bm J} = {\bm j}_{\Lambda^{+}_{c}} 
+ {\bm j}_n$ and  ${\bm J}' = {\bm  j}_p + {\bm j}_\Lambda$ in ${\sf M}^K$. 
As~${\sDelta}^M$ is rotationally invariant, it turns out 
that $J = J'$, which leads to
\beq
{\sf M}^\pi- {\sf M}^K
= \sum_{JM} (j_\Lambda m_\Lambda j_pm_p|JM)\, (j_{\Lambda^{+}_{c}} m_{\Lambda^{+}_{c}} j_n m_n|JM) 
({\sf M}^\pi_J-(-)^{j_\Lambda+ j_p+J}{\sf M}^K_J),
\label{36} 
\eeq
where the phase $(-)^{j_\Lambda+ j_p+J}$ comes from the property of Clebsh-Gordan coefficients
$(j_p m_pj_\Lambda m_\Lambda |JM)=(-)^{j_\Lambda+ j_p+J} (j_\Lambda m_\Lambda j_p m_p|JM) $.
In more detail, one has for the pion contribution
\bea
\hspace{-2.5cm}
{\sf M}^\pi_J&=& -i\int d\xb\, d\yb{\sDelta}^\pi(|\xb-\yb|) 
\left\{ \left[
\A^\pi(t_\Lambda,t_p) \rho_A(\xb) -i \B^\pi(t_\Lambda,t_p) \rho_B(\xb) \right]
 \rho_C(\yb) \right\}_{(J)} ,
\label{37}
\eea
where $(J)\equiv{(j_\Lambda j_p,j_{\Lambda^{+}_{c}} j_n;J)}$ indicates the 
angular momentum coupling described above, and the transition densities are
\bea
\rho_A(\xb) &=& {\psip}^*_{p_\Lambda\kappa_\Lambda}(\xb){{\Psip}_{\kappa_{\Lambda^{+}_{c}} }}(\xb)
- {\psin}^*_{p_\Lambda\kappa_\Lambda}(\xb){{\Psin}_{\kappa_{\Lambda^{+}_{c}}}}(\xb), \label{42a}
\nn\\
\rho_B(\xb) &=& {\psip}^*_{p_\Lambda\kappa_\Lambda}(\xb){{\Psin}_{\kappa_{\Lambda^{+}_{c}} }}(\xb)
+ {\psin}^*_{p_\Lambda\kappa_\Lambda}(\xb){{\Psip}_{\kappa_{\Lambda^{+}_{c}}}}(\xb), 
\nn\label{42b} \\
\rho_C(\yb) &=& {\psip}^*_{p_p\kappa_p}(\yb){{\Psin}_{\kappa_n }}(\yb)
+ {\psin}^*_{p_p\kappa_p}(\yb){{\Psip}_{\kappa_n}}(\yb) .
\label{38}
\eea
To carry out the coordinate integrations it is convenient to perform a tensor expansion of the 
propagators in Eq.~\rf{31}, for $M=\pi,K$, as follows: 
\beq
{\sDelta}^M(|\xb-\yb|) = \sum_{L}{\sDelta}^M_L(x,y) \, [Y_L(\xh) \cdot Y_L(\yh)] ,
\label{39}
\eeq
where $Y_L(\xh)$ denotes the spherical tensor whose components are the spherical harmonics
$Y_{LM}(\xh)$ and similarly for $Y_L(\xh)$, and
\beq
{\sDelta}^M_{L}(x,y)=2\pi \int d(\cos\theta_{xy}) \, {\sDelta}^M(|
\xb-\yb|) \, P_L(\cos\theta_{xy}),
\label{40}
\eeq
with $P_L$ being a Legendre polynomial. Thus, Eq.~\rf{37}  becomes 
\br
{\sf M}^{\pi}_J&=& -i\sum_{L}\int d\xb\, d\yb {\sDelta}^{\pi}_L(x,y)[Y_L(\xh) \cdot Y_L(\yh)] 
\nn\\
&\x&\left\{ \left[
\A^{\pi}(t_\Lambda,t_p) \rho_A(\xb) -i \B^{\pi}(t_\Lambda,t_p) \rho_B(\xb) \right]
 \rho_C(\yb) \right\}_{(J)},
\label{41}
\er
Making use of the well known property of the scalar product of two tensor operators~\cite{de63}
 \beq
\bra{j_1j_2 J}[Y_{L}(\xh)\cdot Y_{L}(\yh)]\ket{j_3j_4J}\;=\; (-)^{j_2+j_3+J}
\sixj{j_1}{j_2}{J}{j_4}{j_3}{L}\Bra{j_1}Y_L\Ket{j_3}\Bra{j_2}Y_L\Ket{j_4},
\label{42}
\eeq
and defining
\bea
A^L_{\kappa\kappa_{\Lambda^{+}_{c}}}(rp) &=&
[f_{p\kappa}(r)F_{\kappa_{\Lambda^{+}_{c}}}(r)-g_{p\kappa}(r)G_{\kappa_{\Lambda^{+}_{c}}}(r)] \,
\Bra{\kappa}Y_{L}\Ket{\kappa_{\Lambda^{+}_{c}}},
\nn\label{ABCa}\\
B^L_{\kappa\kappa_{\Lambda^{+}_{c}}}(rp) &=&
[f_{p\kappa}(r)G_{\kappa_{\Lambda^{+}_{c}}}(r)+g_{p\kappa}(r)F_{\kappa_{\Lambda^{+}_{c}}}(r)]
\, \Bra{-\kappa}Y_{L}\Ket{\kappa_{\Lambda^{+}_{c}}},
\nn\label{ABCb}\\
C^L_{\kappa\kappa_n}(rp) &=&
[f_{p\kappa}(r)G_{\kappa_n}(r)+g_{p\kappa}(r)F_{\kappa_n}(r)]
\, \Bra{-\kappa}Y_{L}\Ket{\kappa_n},
\label{43}
\eea
a trivial, but tedious algebra gives 
\beq
{\sf M}_J^{\pi} = \sum_{L}(-)^{j_p+j_{{\Lambda^{+}_{c}}}+J}
\sixj{j_\Lambda}{j_p}{J}{j_{n}}{j_{\Lambda^{+}_{c}}}{L}{\sf M}_L^{\pi},
\label{44}
\eeq
where
\br
{\sf M}_L^{\pi} &\equiv&-
\int dx dy\, x y \; [\B^{\pi}(t_\Lambda,t_p) B^L_{\kappa_\Lambda\kappa_{\Lambda^{+}_{c}}}(xp_\Lambda)
+i\A^{\pi}(t_\Lambda,t_p) A^L_{\kappa_\Lambda\kappa_{\Lambda^{+}_{c}}}(xp_\Lambda)] \,\nn\\
&\x&{\sDelta}^{\pi}_{L}(x,y) \, C^L_{\kappa_p\kappa_n}(yp_p) .
\label{45}
\er
In the convention adopted in Eq.~(\ref{13}), the reduced matrix elements 
to be used in Eq.~\rf{43} are%
\footnote{The phases appearing in the corresponding equations in Ref.~\cite{Fontoura:2015kta}
are for the opposite ordering in the spin-orbit coupling. This is innocuous for the
rates, but may be important for other observables. Notice also that, irrespectively
of the 
spin-orbit ordering, one has~\cite{Boh69}:
$\Bra{\kappa}Y_L\Ket{\kappa'} = (-)^{\kappa+\kappa'}\,\Bra{\kappa'}Y_L\Ket{\kappa}$.}
\bea
\hspace{-2.1cm}
\Bra{\kappa}Y_L\Ket{\kappa'} &=& (4\pi)^{-1/2}(-)^{l+l'+j'-1/2}\hat{j}\hat{j}'\hat{L}
\threej{j}{L}{j'}{-\fot}{0}{\fot}\frac{1+(-)^{l+l'+L}}{2} ,
\nn\\
\hspace{-2.1cm}
\Bra{-\kappa}Y_L\Ket{\kappa'} &=& (4\pi)^{-1/2}(-)^{\bar{l}+l'+j'-1/2}\hat{j}\hat{j}'\hat{L}
\threej{j}{L}{j'}{-\fot}{0}{\fot}\frac{1+(-)^{{\bar l}+l'+L}}{2},
\label{46}
\eea
which satisfy the following symmetry property:  $\Bra{\kappa}Y_L\Ket{{-\kappa'}}
=\Bra{-\kappa}Y_L\Ket{\kappa'}$.
The amplitude for the kaon contribution can be easily obtained by 
making the following substitutions in Eqs.~\rf{44} and \rf{45}:  
${\sf M}_J^{\pi}, {\sf M}_L^{\pi} \to {\sf M}_J^K, {\sf M}_L^K$,
$j_{\Lambda},\kappa_{\Lambda},p_{\Lambda} \leftrightarrow  j_{p},\kappa_{p},p_{p}$, 
$\A^{\pi}(t_\Lambda,t_p),\B^{\pi}(t_\Lambda,t_p) \rightarrow 
\A^{K}(t_p,t_\Lambda),\B^{K}(t_p,t_\Lambda)$, 
and 
$\sDelta^\pi,{\sDelta}^\pi_L \rightarrow\sDelta^K,{\sDelta}^K_L$.

\section{Integration over $T_p$ in Eq.~(\ref{66a})}
\label{C}

The first step towards the evaluation of the integration over $T_p$, we use
Eq.~(\ref{TR}) so that $f(T_p)$ can be written as
\be
f(T_{p}) = a + T_{p} + \sqrt{b + T_{p}(c+T_{p}) + d\,\sqrt{T_{p}(c+T_{p})}} ,
\label{fTp}
\ee
with 
\be
a = T_{\Lambda}-M_{R}-\Delta_{j_{n}}, \hspace{.2cm}
b = M^{2}_{R}+p^{2}_{\Lambda}, \hspace{.2cm}
c = 2M_p, \hspace{.2cm}
d = 2p_{\Lambda}\cos\theta_{\Lambda p}.
\label{abcd}
\ee
We are then faced with an integration of the form
\be
I = \int\, dx\, F(x)\, \delta[f(x)]
\label{I1}
\ee
where $f(x)$ is defined in Eq.~(\ref{fTp}), with $x = T_p$:   
\bea
f(x) &=& a + x + \sqrt{b + x(c+x) + d\,\sqrt{x(c+x)}},
\label{If}
\eea
and $a$, $b$, $c$ and $d$ defined in Eq.~(\ref{abcd}). To eliminate the delta-function, 
we use the identity
\be
\delta[f(x)] = \sum_n \frac{\delta(x-x_n)}{|f'(x_n)|}, 
\label{identity}
\ee
where the summation is over all the real-valued simple zeros of $f(x)$, \ie
\be
f(x_n) = 0, \hspace{.2cm}
f'(x_n) \ne 0.
\label{simple}
\ee
Introducing \Ref{identity} into Eq.~\Ref{I1} gives 
\be
I = {\sum_n}' \frac{F(x_n)}{|f'(x_n)|},
\label{I2}
\ee
where the prime in the summation sign is to remind that only those zeros that fall within the 
region of integration in Eq.~\Ref{I1} are to be included. In our case, $x$ stands for a kinetic energy, 
therefore we must require that $x_n > 0$.

To find the zeros, we need to solve the equation $f(x)=0$, \ie
\bea
\sqrt{b + x(c+x) + d\,\sqrt{x(c+x)}} &=&  - (a + x).
\label{zeroes}
\eea
Squaring it, gives
\bea
d\,\sqrt{x(c+x)}  &=&  (a + x)^2 - [b + x(c+x)],
\label{square1a}
\eea
and squaring this latter expression, gives
\bea
d^2\,x(c+x)  &=& \{ (a + x)^2 - [b + x(c+x)] \}^2.
\label{square2}
\eea
Noticing that
\be
(a + x)^2 - [b + x(c+x)] \equiv (a^2-b) - (c-2a)x,
\label{reduction}
\ee
it is clear that Eq.~\Ref{square2} can be written as
\bea
\alpha\,x^2 + \beta\,x + \gamma &=& 0,
\label{standardd}
\eea
with the coefficients $\alpha,\beta,\gamma$ given in terms of $a,b,c,d$ as
\bea
\alpha &=& (c-2a)^2 - d^2,
\label{aalpha}
\\
\beta &=& -[ 2(a^2-b)(c-2a) + cd^2 ],
\label{bbeta}
\\
\gamma &=& (a^2-b)^2.
\label{ggamma}
\eea
The roots of the quadratic equation \Ref{standardd} are given by
\be
x^\pm_n = \frac{-\beta \pm \sqrt{\beta^2 - 4\alpha\gamma}}{2\alpha} ,
\label{roots}
\ee
with the discriminant given by
\be
\beta^2 - 4\alpha\gamma =
d^2\{ 4(a^2-b) [(c-2a)c + (a^2-b)] + c^2d^2 \}.
\label{discriminant}
\ee

It is important to note that in the manipulations to arrive at Eq.~\Ref{standardd},
spurious solutions might have been introduced and one needs to verify whether these 
two roots do indeed satisfy the original equation, Eq.~\Ref{zeroes}. Only then can 
they be taken as legitimate solutions to our problem. In particular, the 
derivative $f'(x_n)$ is given, at the legitimate zeros, by 
\begin{equation}
f'(x_n ) = 1- \frac{c+ 2x_n}{2(a+x_n)}  -  \frac{d(c+2x_n)}{4  (a+x_n) \sqrt{x_n (c+x_n )} }.
\label{f'x}
\end{equation}

Since $x = T_p$ and $x^\pm = T^\pm_p$, one then have
\be
\delta[f(T_{p})]=\frac{\delta(T_{p}-T_{p}^{+})}{|f'(T^{+}_{p})|}+
\frac{\delta(T_{p}-T_{p}^{-})}{|f'(T^{-}_{p})|},
\label{deltafunc}
\ee
where 
\be
f'(T^{\pm}_{p}) = 1- \frac{M_p+ T^{\pm}_{p}}{T_{\Lambda}-M_{R}
-\Delta_{j_{n}}+T^{\pm}_{p}} - \frac{p_{\Lambda}\cos\theta_{\Lambda p}
(M_p+T^{\pm}_{p})}{(T_{\Lambda}-M_{R}-\Delta_{j_{n}}+T^{\pm}_{p}) 
\sqrt{T^{\pm}_{p} (2M_p+T^{\pm}_{p} )}},
\label{77}
\ee
with
\bea
T_{p}^\pm &=& 
\frac{K_1(T_\Lambda,\cos\theta_{\Lambda p}) \pm 
|p_\Lambda\cos\theta_{\Lambda p}| \sqrt{K_2(T_\Lambda,\cos\theta_{\Lambda p}})}
{2K_3(T_\Lambda,\cos\theta_{\Lambda p})},
\label{Tp-root}
\eea
where 
\bea
\hspace{-1.0cm}
K_1(T_\Lambda,\cos\theta_{\Lambda p}) &=& (M_p+M_R+\Delta_{j_n}-T_\Lambda)\,
[\Delta_{j_n}(2M_R+\Delta_{j_n}) \nn \\
&-& 2(M_\Lambda+M_R+\Delta_{j_n})T_\Lambda] \nn \\
&+& 2M_p\, T_\Lambda(2M_\Lambda+T_\Lambda) \cos^2\theta_{\Lambda p},
\label{K1} \\[0.1true cm]
\hspace{-1.0cm}
K_2(T_\Lambda,\cos\theta_{\Lambda p}) &=& [\Delta_{j_n}(2M_R+\Delta_{j_n}) 
- 2(M_\Lambda+M_R+\Delta_{j_n})T_\Lambda]
\nonumber\\
&\times& [4M_p(M_p+M_R) + \Delta_{j_n}(2M_R+4M_p+\Delta_{j_n})
\nonumber\\
&-& 2(M_\Lambda+M_R+2M_p+\Delta_{j_n})T_\Lambda] \nn \\
&+& 4M_p^2\, T_\Lambda(2M_\Lambda+T_\Lambda) \cos^2\theta_{\Lambda p},
\label{K2} \\[0.1true cm]
\hspace{-1.0cm}
K_3(T_\Lambda,\cos\theta_{\Lambda p}) &=& (M_p+M_R+\Delta_{j_n}-T_\Lambda)^2 - 
T_\Lambda(2M_\Lambda+T_\Lambda) \cos^2\theta_{\Lambda p}.
\label{K3}
\eea

\section*{References}
\label{references}

\end{document}